# Interlayer coupling rotatable magnetic easy-axis in MnSe$_2$ mono- and bi-layers

Zhongqin Zhang[1,2,3,†], Cong Wang[2,3,†,*], Peng-Jie Guo[2,3], Linwei Zhou[2,3], Yuhao Pan[2,3,4,5], Zhixin Hu[1,*] and Wei Ji[2,3,*]

[1]*Center for Joint Quantum Studies and School of Physics, Tianjin University, Tianjin, China*
[2]*Beijing Key Laboratory of Optoelectronic Functional Materials & Micro-Nano Devices, School of Physics, Renmin University of China, Beijing 100872, China*
[3]*Key Laboratory of Quantum State Construction and Manipulation (Ministry of Education), Renmin University of China, Beijing 100872, China*
[4]*China North Artificial Intelligence & Innovation Research Institute, Beijing 100872, China*
[5]*Collective Intelligence & Collaboration Laboratory, Beijing 100872, China*

*Emails: wcphys@ruc.edu.cn (C.W), zhixin.hu@tju.edu.cn (Z.H.), wji@ruc.edu.cn (W.J.)

Interlayer coupling plays a critical role in tuning the electronic structures and magnetic ground states of two-dimensional materials, influenced by the number of layers, interlayer distances, and stacking order. However, its effect on the orientation of the magnetic easy axis remains underexplored. In this study, we demonstrate that interlayer coupling can significantly alter the magnetic easy-axis orientation, as shown by the magnetic easy-axis of monolayer 1$T$-MnSe$_2$ tilting 67° from the $z$-axis, while aligning with the $z$-axis in the bilayer. This change results from variations in orbital occupations near the Fermi level, particularly involving nonmetallic Se atoms. Contrary to the traditional focus on magnetic metal atoms, our findings reveal that Se orbitals play a key role in influencing the easy-axis orientation and topological Chern numbers. Furthermore, we validated our conclusions by changing stacking orders, introducing charge doping, applying in-plane biaxial strains, and substituting non-metallic atoms. Our results highlight the pivotal role of interlayer coupling in tuning the magnetic properties of layered materials, with important implications for spintronic applications.

# I. INTRODUCTION

Two-dimensional (2D) van der Waals (vdW) magnetic materials have garnered significant attention and have achieved notable advancements in recent years [1,2], emerging as a promising platform for both fundamental research [3,4] and potential spintronics applications [5,6]. The magnetic properties of these materials are typically characterized by magnetization [7–9], magnetic orders [2,10,11] and magnetic anisotropy [10,12,13], with the latter two being easily tunable. Extensive discussions have been conducted on both in-plane and out-of-plane magnetic orders, associated net magnetization values, and their manipulation mechanisms for 2D magnets. In addition to these properties, magnetic anisotropy plays a pivotal role in sustaining long-range magnetic orders in 2D magnets at finite temperatures [3,4]. It is also closely associated with magnetic coercivity, a critical parameter determining whether the material behaves as a hard or soft magnet [14–16]. However, results on the magnetic easy-axis are, unlike the magnetic order, available only for a limited number of 2D magnets, such as $CrTe_2$ [10]. This limitation in understanding hinders us to grasp the factors that could affect magnetic anisotropy, thereby limiting the development of effective strategies for modulating magnetic anisotropy.

The magnetic anisotropy was demonstrated to be tunable under various external stimuli, such as electric fields [17–19] and charge doping [20,21], which primarily change the filling of *d* orbitals of magnetic metal atoms. However, such tunability typically requires high stimulating strengths, which can lead to magnetic order transitions [22,23] or irreversible structural phase transitions [24,25]. Thus, it is of paramount importance to explore gentle and sustainable mechanisms for rotating the magnetic easy axis direction without introducing transitions of magnetic order or atomic structures. The ability to modulate various properties through interlayer coupling is one of the most striking features of 2D materials [19]. In 2D magnets, many demonstrations highlight the role of interlayer coupling in manipulating interlayer or intralayer magnetic order [28–32]. Although the orientation of the magnetic easy-axis was observed to vary upon changing the number of layers in 1*T*-$CrTe_2$ [33,34] and

alpha-RuCl$_3$ [35]. However, these systems also exhibit synchronous magnetic order transition [33,34] or structural symmetry-breaking [35] with layer number variation. The question of whether interlayer coupling can tune magnetic anisotropy independently remains largely unexplored.

In this article, we theoretically explored the ability and mechanism to tune the magnetic easy axis in prototypic MnSe$_2$ mono- and bi-layers via interlayer coupling using density functional theory (DFT) calculations. We found that the magnetic easy-axis rotates from 67º off the *z*-axis in the monolayer to the *z*-axis in the bilayer, without any transition of magnetic orders or atomic structures. By analyzing the contribution of each individual atom and orbital to the MAE, we identified that the electron occupation of Se *p* orbitals is critical in driving this rotation. A comparison of the electronic band structures of the mono- and bi-layers reveals the mechanism behind the occupation changes and the subsequent magnetic easy-axis rotation through interlayer electronic couplings upon stacking. This mechanism was further validated by tuning the occupation using other external stimuli such as stacking order changes, electron or hole doping, in-plane biaxial strains and substitution of non-metallic atoms. Additionally, we observed that the Chern number of electronic bands near the Fermi level varies in response to the rotating magnetic moments and/or layer stacking, demonstrating ability to manipulate topological properties of these electronic states through various external stimuli.

## II. METHODS

Our density functional theory (DFT) calculations were performed using the generalized gradient approximation (GGA) for the exchange correlation potential in the form of PerdewBurke–Ernzerhof (PBE) [39], the projector augmented wave method [40], and a plane-wave basis set as implemented in the Vienna ab-initio simulation package (VASP) [41]. We also included the dispersion correction through Grimme's semiempirical D3 scheme [42] in combination with the PBE functional (PBE-D3). This combination yields accuracy comparable to that of the optB86b-vdW functional for describing geometric properties of layered materials(Appendix A) [43], but at a lower computational cost. The MAE in MnSe$_2$ is on the order of 0.1 meV, and during the relaxation of the volume and lattice shape, the Pulay stress issue was also encountered. Therefore, to accurately compute the atomic structures, electronic structure, and MAE, kinetic energy cutoffs of 800 and 600 eV for the planewave basis were adopted for structural relaxations and electronic structure calculations, respectively. All atoms, lattice volumes, and shapes were allowed to relax until the residual force per atom was below 0.01 eV/Å. A vacuum layer exceeding 15 Å in thickness (22 Å for the monolayers and 18 Å for the bilayers) was employed to reduce imaging interactions between adjacent supercells. A Gamma-centered *k*-mesh of 21×21×1 was used to sample the first Brillouin zone of the unit cell for MnSe$_2$. The Gaussian smearing method with a $\sigma$ value of 0.01 eV was applied for structural relaxation and electronic structure calculations. To accurately calculate the MAE, the tetrahedron method with Blöchl corrections and a Gamma-centered k-mesh of 29×29×1 was used. The on-site Coulomb interaction for Mn *d* orbitals was characterized by *U* and *J* values of 4.0 eV and 0.7 eV [44], respectively, as determined via a linear response method [45] and validated by the convergence of theoretical predictions(Appendix A). A $2\times 2\sqrt{3}$ supercell and four (eight) magnetic configurations (Appendix B) were considered to identify the magnetic ground state for the monolayer (bilayer). Spin-exchange coupling parameters were extracted based on an anisotropic nearest-neighbor Heisenberg model; please refer to Appendix C for details.

The Atomic-orbital Based Ab-initio Computation at USTC (ABACUS) package [46,47] and PYATB [48] were used to calculate the Chern numbers. By using the Wannier90 package, we constructed the tight-binding model of MnSe$_2$ with Mn 3*d* and Se 4*p* orbitals based on the maximally localized Wannier functions method (MLWF) [49]. We further plotted edge states and verified our conclusions regarding topological transitions using the WannierTools software package [50].

Direct charge doping was applied to Se atoms using the ionic potential method [51]. Specifically, electrons (or holes) were extracted from a 3*d* core level of Se and placed into the lowest unoccupied band of MnSe$_2$. This method ensures that the doped charges remain localized around the Se atoms while maintaining the neutrality of the layer. In-plane biaxial strain was simulated on the monolayer MnSe$_2$, which was fully relaxed in the ferromagnetic configuration. When biaxial strains were applied, the in-plane lattice vectors of the equilibrium configuration were compressed or stretched. The ferromagnetic configuration was used for structural relaxations and its superior stability was further verified by comparing its energy with another three magnetic configurations. For example, to model a -2% biaxial strain, the equilibrium in-plane lattice vectors were compressed from 3.61 Å to 3.54 Å(by 2%). In subsequent structural relaxations, all atoms are were allowed to move and the in-plane lattice constants are were kept fixed.

The spin-orbit coupling was considered in all calculations carried out to determine the magnetic easy-axis directions. The magnetic easy-axis was identified using the Renmin Magnetic Easy Axis Finder (ReMEAF) toolkit [52], which utilizes the simulated annealing algorithm and invokes VASP to determine the global easy-axis orientation. In angular dependence of MAE, for the area between calculated points, we used linear interpolation to give MAE values.

Spin-exchange coupling parameters were extracted based on Heisenberg model as follow:

$$H = H_0 - J_1 \sum_{\langle i \neq j \rangle} \vec{S_i} \cdot \vec{S_j} - J_2 \sum_{\langle\langle i \neq j \rangle\rangle} \vec{S_i} \cdot \vec{S_j} - J_3 \sum_{\langle\langle\langle i \neq j \rangle\rangle\rangle} \vec{S_i} \cdot \vec{S_j}$$

$$-J_4 \sum_{\langle i,j \rangle} \vec{S_i} \cdot \vec{S_j} - J_5 \sum_{\langle\langle i,j \rangle\rangle} \vec{S_i} \cdot \vec{S_j} - J_6 \sum_{\langle\langle\langle i,j \rangle\rangle\rangle} \vec{S_i} \cdot \vec{S_j}$$

Here, $J_1$~$J_3$ and $J_4$~$J_6$ represent the three nearest intra- and interlayer couplings, respectively, as illustrated in Fig. 1(a)~(c). Here, the method for extracting the exchange constants $J_1$, $J_2$, and $J_3$ in the monolayer is presented, and that for the bilayer can be found in Appendix B. First, we wrote the expressions of the Heisenberg model for different magnetic configurations shown in Fig. 6 a-d:

$$E_a = -\frac{N^2}{4} \times \frac{1}{2}(6J_1 + 6J_2 + 6J_3) + E_0$$

$$E_b = -\frac{N^2}{4} \times \frac{1}{2}(2J_1 - 2J_2 - 2J_3) + E_0$$

$$E_c = -\frac{N^2}{4} \times \frac{1}{2}(-2J_1 - 2J_2 + 6J_3) + E_0$$

$$E_d = -\frac{N^2}{4} \times \frac{1}{2}(-2J_1 + 2J_2 - 2J_3) + E_0$$

where $N$ represents the unpaired spins on each Mn atom, which is treated as 3 in our exchange parameter calculations and $E_0$ represents the energy that is independent of the magnetic configuration. Then, by substituting the energies of the magnetic configurations from Table 5, we solved for the values of the exchange constants.

## III. RESULTS AND DISCUSSIONS

### A. Magnetic anisotropy in MnSe$_2$ monolayer and bilayer

Monolayer MnSe$_2$ crystallizes in a hexagonal 1$T$ structure with space group *P-3M1*, as depicted in Fig. 1a. Each Mn atom is octahedrally coordinated by six Se atoms, with Se-Mn-Se bond angle close to 90° (measured as either 90.96° or 89.04°), indicating minimal Jahn-Teller distortion. In the MnSe$_2$ bilayer, the AA stacking (Fig. 1b and 1c) is found to be over 4 meV/Mn more stable compared to other stacking configurations (Appendix D). To determine intra- and interlayer magnetic ground state, we adopted same structure using the optimized structures with the ferromagnetic (FM) order and calculated relative energies of four (eight) magnetic configurations for monolayer (bilayer) MnSe$_2$ (see Appendix B for more details). We found that both the monolayer and the bilayer exhibit a FM ground state, as indicated by the data listed in Table 5, consistent with the theoretical predictions reported in the literature [28,53].

Three intra-layer ($J_1$ to $J_3$) and three inter-layer ($J_4$ to $J_6$) spin-exchange coupling parameters, which are indicated by dashed arrows in different colors in Fig. 1a-1c, were computed based on the Heisenberg model (seen in II. METHODS and Appendix B), , with their values provided in Table 1. Notably, the dominant and positive nearest-neighbor intralayer spin-exchange coupling $J_1$ (7.5 and 8.7 meV/Mn for the mono- and bi-layer) aligns with the energetically favored intralayer FM configuration. Comparable values for the interlayer spin-exchanges, $J_4$=3.6 meV/Mn and $J_5$=3.5 meV/Mn, indicate that the interlayer FM coupling is as strong as the intra-layer magnetic coupling in MnSe$_2$, which is prominent in layered magnetic materials [29,43,54].

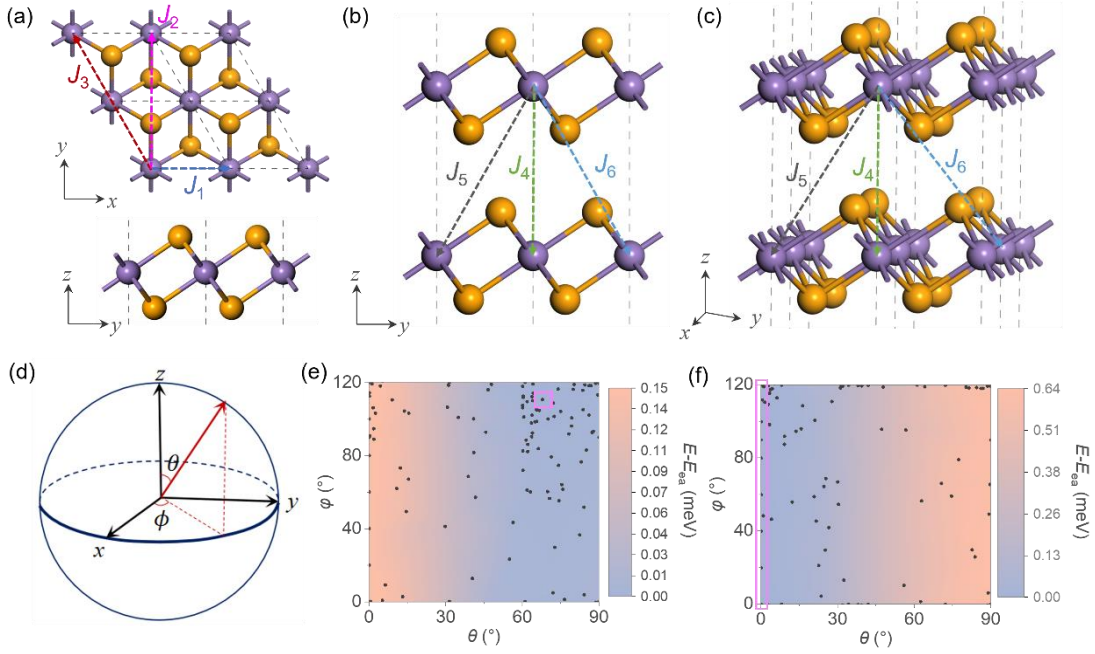

Fig. 1. Structure and easy axis of mono- and bilayer MnSe$_2$. (a) Top and side views of monolayer MnSe$_2$. Purple and orange balls represent Mn and Se atoms, respectively. Dashed arrows denote intralayer spin-exchange parameters $J_1$, $J_2$ and $J_3$ between Mn sites with different colors. (b-c) Side and oblique view of an AA-stacked bilayer MnSe$_2$. Colored dashed arrows denote interlayer spin-exchange parameters $J_4$, $J_5$ and $J_6$ between Mn sites. (d) Definition of polar angle $\theta$ and azimuth angle $\varphi$ in the spherical coordinate. (e-f) Angular dependence of the calculated MAE of monolayer(e) and bilayer(f) MnSe$_2$. The total energy of the Mn moment oriented to the direction of easy axis was chosen as the zero-energy reference and marked by pink boxes. Black dots represent MAE results for angles calculated with DFT.

Table 1. Lattice constant, magnetic ground state, exchange parameters and easy axis of MnSe$_2$. Magnetic Anisotropy Energy (MAE) is defined as the energy difference required to reorient the magnetic moment of MnSe$_2$ from its easy axis to hard axis.

| Layer Number | Lattice Constant (Å) | Mag. Config. | | Exchange Parameters (meV/Mn) | | | | | | Easy Axis | | MAE (meV/Mn) |
|---|---|---|---|---|---|---|---|---|---|---|---|---|
| | | Intralayer | Interlayer | $J_1$ | $J_2$ | $J_3$ | $J_4$ | $J_5$ | $J_6$ | $\theta(°)$ | $\varphi(°)$ | |
| 1L | 3.61 | FM | - | 7.5 | 1.2 | 0.2 | - | - | - | 67 | 111 | 0.2 |
| 2L | 3.63 | FM | FM | 8.7 | 0.3 | 1.0 | 3.6 | 3.5 | 0.7 | 0 | - | 0.6 |

By considering the $C_{3v}$ spatial symmetries inherent in MnSe$_2$, a range of polar angle $\theta \in [0, 90°]$ and azimuth angle $\varphi \in [0, 120°]$ (Fig. 1d) covers all possible magnetic easy-axis orientations. Figure 1e plots the magnetic anisotropic energies of the MnSe$_2$ monolayer where its magnetization direction rotates as a function of $\theta$ and $\varphi$, revealing an easy axis orientation along $\theta=67°$ and $\varphi=111°$. When an additional layer is stacked onto the monolayer to form the bilayer, the easy axis rotates to align with the $z$-axis ($\theta=0$, Fig. 1f). Such a substantial rotation of the magnetic easy-axis direction by adding a single layer to a monolayer has yet to be reported in 2D layered magnetic materials and warrants further exploration.

### B. Origin of the interlayer coupling tunable magnetic easy-axis direction

Figure 2a shows the interlayer differential charge density of the bilayer MnSe$_2$. Apparent charge reduction (blue contours) near the interlayer Se atoms (pink balls) and charge accumulation mainly in the vdW gap region (light yellow contours) suggest strong interlayer electronic hybridization. Meanwhile, charge transfer on Mn and surface Se (orange balls) is less significant, indicating a weaker effect of interlayer coupling on them and thus a different contribution to the easy-axis rotation. To understand the role of interlayer coupling in rotating the easy axis, we decomposed the MAE contribution into individual atoms (Fig. 2b). Here, positive and negative MAE contributions correspond to magnetic easy-axis directions tending towards the $z$-axis and in-plane directions, respectively. In the MnSe$_2$ monolayer, all Se atoms are categorized as Se-surface atoms and contribute a positive value (0.36 meV/atom) to the MAE, while Mn atoms donate a smaller negative contribution (-0.04 meV/atom) to the MAE, favoring an in-plane magnetic easy-axis. The competition between these contributions results in the tilted orientation of the magnetic easy-axis in the monolayer. After the stacking of an additional MnSe$_2$ layer, the MAE contributions for Mn and surface Se atoms change sign but have smaller absolute values (no more than 0.25 meV/atom). However, the Se-interface atoms exhibit a dominant positive MAE contribution (1.58 meV/atom), decisively outweighing the negative contribution from Se-surface (-0.25 meV/atom), leading to the alignment of the magnetic easy axis with the $z$-axis in the bilayer.

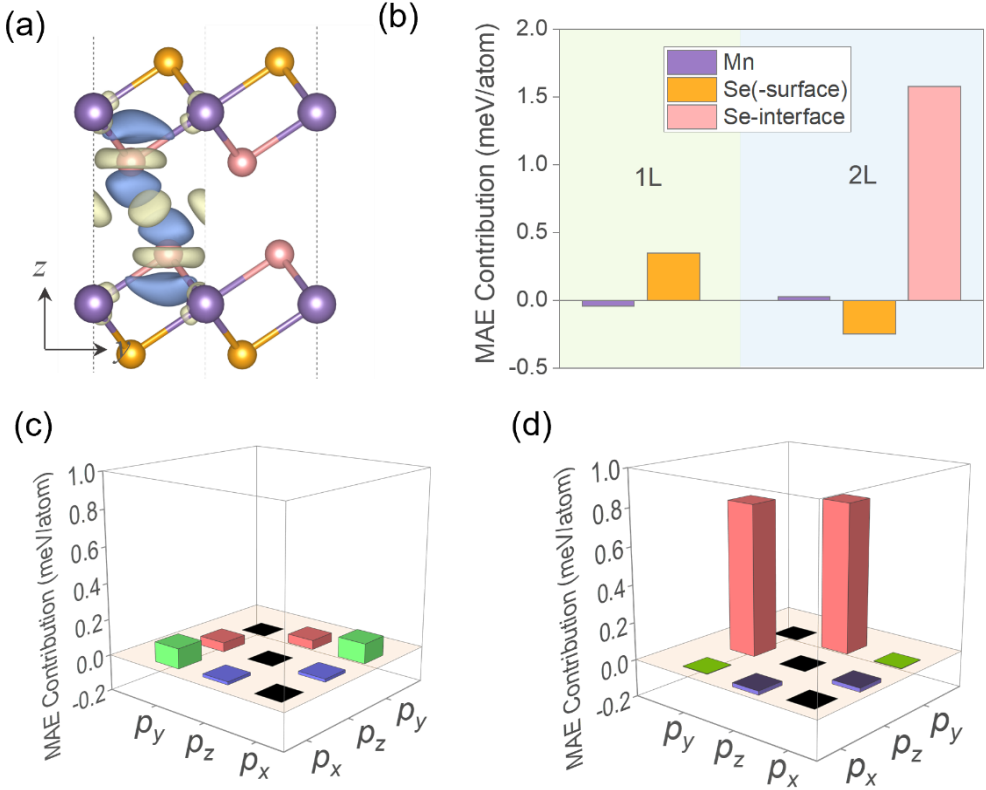

Fig. 2 (a) Side view of interlayer differential charge density (DCD) of bilayer MnSe$_2$ with an isosurface value of 0.0005 $e$/Bohr$^3$. Light yellow and blue isosurface indicates charge accumulation and reduction after layer stacking. Non-equivalent Se atoms in bilayer MnSe$_2$ are colored in orange and pink. (b) Atomically decomposed MAE contributions of monolayer(1L) and bilayer(2L) MnSe$_2$. (c-d) Orbital-resolved MAE contributions of Se in monolayer(c) and Se-interface in bilayer(d).

Therefore, we further focused on the origin of the different Se contributions to MAE across different numbers of layers and decomposed the MAE into Se $p_x$, $p_y$, and $p_z$ orbitals for the mono- (Fig. 2c) and bi-layer (Fig. 2d), namely:

$$\langle p_i|H_{SOC}(x)|p_j\rangle - \langle p_i|H_{SOC}(z)|p_j\rangle$$

where $H_{SOC}(x)$ is the spin-orbit coupling Hamiltonian when the magnetic moment is oriented along the *x*-direction, and the $p_i$, $p_j$ represents $p_x$, $p_y$ or $p_z$. In the monolayer, all Se orbital contributions are relatively small , resulting in the moderate total contribution of 0.36 meV/atom. As for the bilayer, the contribution from the $p_z$-$p_y$ component of Se-interface atoms (~0.8 meV/atom) is significantly prominent and at least an order of magnitude larger than other components (less than 0.03 meV/atom). This result indicates that the interaction between the $p_z$ and $p_y$ orbitals of the Se-interface atoms is crucial in orienting the magnetic easy-axis toward the normal direction in the MnSe$_2$ bilayer. This finding provides a different perspective from the prevailing view in the literature, which attributes easy-axis reorientation due to interlayer couplings primarily to the influence of metallic atoms.

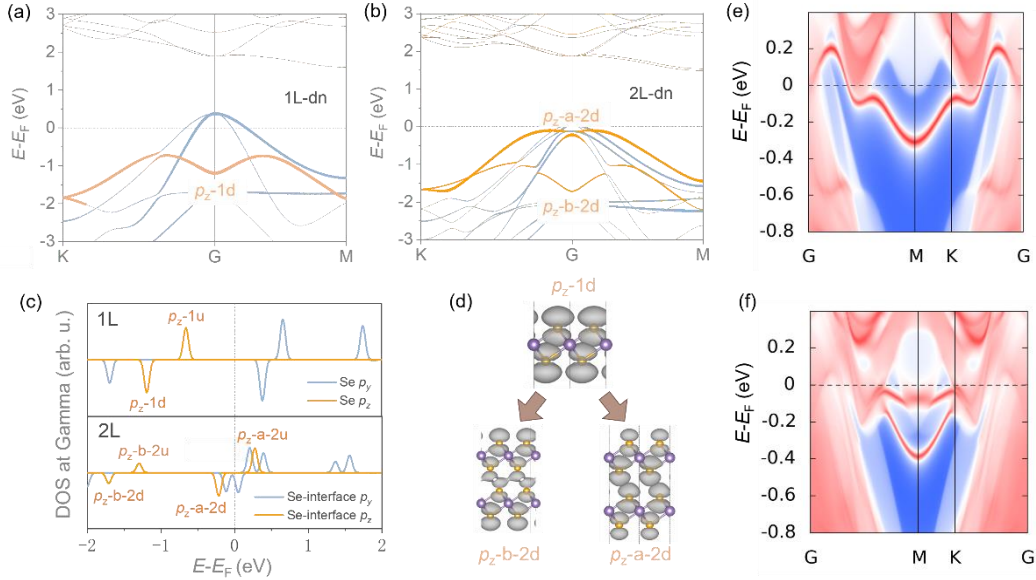

Fig. 3 Electronic structures of MnSe$_2$. (a-b) Spin-down band structures of monolayer(a) and bilayer(b) MnSe$_2$. The $p_z$ and $p_y$ orbitals of (interfacial) Se are mapped with different colors in bands near the Fermi level: orange, $p_z$; light blue, $p_y$. The Fermi level is marked using grey dot dash line. (c) Projected density of states for the $p_y$ and $p_z$ orbitals of (interfacial) Se at the Gamma point in monolayer (upper panel) and bilayer (lower panel). (d) Visualized wave-function norms for the labeled states in (a-c). The isosurface value is 0.0012 e/Bohr$^3$. (e,f) Surface states of monolayer(e) and bilayer(f) MnSe$_2$.

To understand the origin of the pronounced $p_z$-$p_y$ component, we mapped the orbitals of (interfacial) Se on the band structures of both monolayer and bilayer. The orbital decomposition and layer number dependence of electronic band structures show qualitative consistency with different spin components. For clarity, we examine the spin-down electronic structure here (Fig.3a-b) and provide the details of spin-up results in the Appendix G. According to second-order perturbation theory [55], the states with different occupations near the Fermi level contribute most significantly to the MAE, primarily composed of $p_z$ and $p_y$ orbitals in both mono- and bilayer $MnSe_2$. For a clear comparison, we plot the projected density of states (PDOS) of Se $p$ states at G point in Fig.3c. In the monolayer, the $p_z$ states are distant from the Fermi level, resulting in an energy difference of approximately 0.85eV between the $p_z$ and $p_y$ states with different occupations (Fig. 3a and the upper panel of Fig. 3c). In the bilayer, the interfacial Se $p$ orbitals overlap and hybridize into bonding ($p_z$-b-2d) and antibonding ($p_z$-a-2d) states to release Pauli/Coulomb repulsions (Fig. 3d), significantly splitting the $p_z$ bands. This reduces the energy differences between $p_z$ and $p_y$ states around Fermi level with different occupation to 0.1eV at G point (Fig. 3b and the lower panel of Fig. 3c), thereby enhancing $p_z$-$p_y$ interactions and increasing MAE contribution favoring the $z$ axis (Fig. 2d).

We also explored the possibility that the differences in Kitaev interactions in the mono- and bi-layer $MnSe_2$ might modulate the magnetic easy-axis direction. However, as listed in Appendix C, the non-collinear spin exchange and Kitaev interactions are at least two orders of magnitude weaker than that of the isotropic spin exchange coupling $J_1$ and remain nearly unchanged with increasing the number of layers, indicating their negligible influence on the magnetic easy-axis direction in $MnSe_2$ mono- and few-layers.

Furthermore, we found that monolayer and bilayer $MnSe_2$ are topologically nontrivial semimetals. We term the three energy bands crossing the Fermi level as band $N$-1, band $N$ and band $N$+1, respectively (Appendix H). Forming interlayer bonding states directly modifies band structures near the Fermi level, while layer-number-induced easy-axis rotation changes the magnetic group from P-1.1 in the monolayer to P-3m'1 in the bilayer, collectively leading to layer-number-tunable Chern numbers and surface states. (Table 2 and Fig. 3(e-f)). Both magnetic anisotropy and topological properties are governed by the bands around the Fermi level, suggesting possible magnetic-field-manipulated topological features. In addition to layer stacking, applying a vertical magnetic field can also reorient the magnetic moments in the monolayer from the easy axis to the $z$-axis, thereby altering the band structures and magnetic group (from P-1.1 to P-3m'1). This shift results in variations in the topological Chern numbers of the three non-trivial bands(Appendix H and Table 2), similar to what is observed in CeX (X=Cl, Br, I) [56].

Table 2 Chern numbers of MnSe$_2$ in 1L and 2L with different numbers of occupied bands and directions of magnetic moments.

| Layer Number | | 1L | | 2L |
|---|---|---|---|---|
| Direction | | Easy Axis | $z$ | Easy Axis($z$) |
| Band No. | $N+1$ | -4 | -2 | 2 |
| | $N$ | 0 | 0 | 0 |
| | $N-1$ | 0 | -4 | -5 |

### C. Modulation of the easy axis direction in MnSe$_2$ monolayer

In addition to the magnetic field, stacking orders, charge doping and external strain can also effectively change the electron band structure around the Fermi level and thus the easy axis direction. Besides the most stable AA stacking, we considered five additional stacking orders and demonstrated that the interlayer stacking can effectively control the magnetic ground state and easy-axis direction (Appendix D) [28]. The AB stacking configuration (Fig. 4(a)) shares the same FM groundstate with AA stacking but exhibits a different in-plane easy-axis. To understand how stacking order rotates the orientation of the easy axis, we plotted the contributions of different atoms to the MAE of both AA and AB stacked bilayers (Fig. 4(b)). In the transition from AA to AB stacking, there is no qualitative change in the favored easy-axis direction of different atoms, but the MAE contribution of the Se-interface decreases significantly by threefold, while the contribution of the Se-surface nearly doubles. As a result, the reduced MAE contribution of the Se-interface becomes less competitive compared to that of the Se-surface, leading to an in-plane easy axis in the AB stacked bilayer. This change in the Se-interface MAE contribution arises from the weakened interaction between the $p_z$ and $p_y$ states (Appendix I).

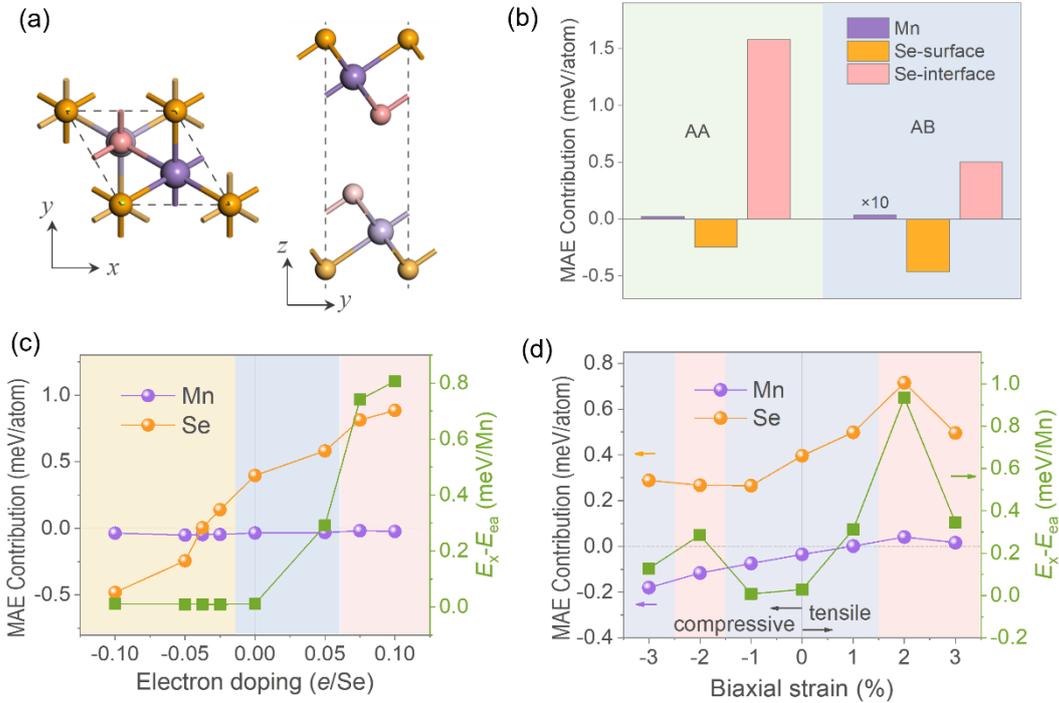

Fig. 4 Modulation of the magnetic anisotropy of MnSe$_2$. (a) Structure model of AB stacked bilayer. (b) Atomically decomposed MAE of AA and AB stacking. (c-d) Atomically decomposed MAE and the energy difference required to reorient the magnetic moments from the easy axis ($E_{ea}$) to the $x$ axis ($E_x$) as a function of doping concentrations(c) and in-plane biaxial strain(d) in monolayer MnSe$_2$. The light brown, blue, and pink shading indicate that the easy axis lies in the $xy$-plane, tilts away from the $z$-axis, and aligns with the $z$-axis, respectively.

Charge doping and in-plane strain are commonly introduced by substrates. They are often employed to manipulate the magnetism of 2D magnets. Therefore, we considered the influence of charge doping and in-plane strain on the magnetic anisotropy of monolayer MnSe$_2$. By varying doping concentrations (Fig. 4c) and applying biaxial strain (Fig. 4d), the orientations of the easy axis of monolayer MnSe$_2$ can be manipulated. When the doping concentration is adjusted to either -0.05 or 0.075 e/Se, or when biaxial strain reaches -2% or 2%, the easy axis undergoes a rotation towards the *xy*-plane or *z*-axis (Appendix I), showcasing the adjustability of the easy axis in monolayer MnSe$_2$. The atomically decomposed MAE in Fig. 4c,d reveals that the MAE contribution of Se exhibits significant variability, whereas Mn consistently makes relatively minor contributions, corroborating earlier findings.

Finally, to further support the importance of non-metallic atoms in orientating the easy axis, we substituted the Se atoms with Te atoms and studied the magnetic anisotropy of monolayer 1T-MnTe$_2$. The 1T-MnTe$_2$ exhibits a large out-of-plane magnetic anisotropy with a MAE value of 1.6 meV/Mn, where the contribution from Te (1.7 meV/atom favoring the *z*-axis) is much larger than that from Mn (-0.3 meV/atom). The difference in the easy-axis and MAE between MnSe$_2$ and MnTe$_2$ again demonstrates the substantial contributions of nonmetallic atoms on tuning magnetic anisotropy in MnX$_2$.

## IV. CONCLUSIONS

In summary, we found that MnSe$_2$ is ferromagnetic topological semimetal, with its easy-axis direction and topological properties highly dependent on the electronic band structure near the Fermi level. By analyzing atomically and orbitally decomposed MAE and electronic structures, we revealed the mechanism by which interlayer coupling changes the direction of the easy-axis from 67º off the *z*-axis in the monolayer to the *z*-axis in the bilayer. In MnSe$_2$, the electronic states near the Fermi level are mainly contributed by non-metallic Se atoms, resulting in the significant influence of interlayer coupling between interfacial Se atoms on the electronic band structure. From the monolayer to bilayer, $p_z$ orbitals of Se are split towards the Fermi level due to interlayer coupling, leading to the pronounced $p_z$-$p_y$ interaction and MAE contribution favoring *z*-axis. Furthermore, based on the modification of Se electronic states, we demonstrated that the orientation of the magnetic easy-axis can also be manipulated by stacking orders, doping, biaxial strain and substitution of non-metallic atoms. Our results advance the understanding of the mechanism behind the rotation of the easy axis orientation in 2D layered magnets.

## ACKNOWLEDGMENTS

We thank Prof. Jun Hu at Ningbo University, Prof. Lixin He and Gan Jin at University of Science and Technology of China for valuable discussions. We gratefully acknowledge the financial support from the Ministry of Science and Technology (MOST) of China (Grant No. 2023YFA1406500), the National Natural Science Foundation of China (Grants No. 92477205, No. 52461160327, and No.12104504), the Fundamental Research Funds for the Central Universities, and the Research Funds of Renmin University of China [Grants No. 22XNKJ30 (W.J.) and 24XNKJ17 (C.W.)]. All calculations for this study were performed at high performance cluster at center for joint quantum studies (HPC-CJQS) of Tianjin University, the Physics Lab of High-Performance Computing (PLHPC) and the Public Computing Cloud (PCC) of Renmin University of China.

## APPENDIX A: EFFECTS OF FUNCTIONAL AND DIFFERENT U-VALUES ON THE GEOMETRY AND MAGNETIC GROUND STATE OF MONOLAYER MnSe$_2$

Table 3 Lattice Constant, magnetic moment per Mn atom and layer thickness $d_l$ of monolayer MnSe$_2$ calculated with different exchange-correlation functionals. To ensure the robustness and applicability of our methods, we evaluated the equilibrium lattice constants of the FM configuration using different functionals, e.g., with or without various forms of vdW correction, and with or without *UJ* correction. We choose the PBE-D3+*UJ* for structure optimization and electronic structures.

| Functional | Lattice Constant(Å) | Mag.Mn($\mu_B$) | $d_l$(Å) |
|---|---|---|---|
| PBE-w/o UJ | 3.48 | 2.89 | 2.87 |
| PBE+UJ | 3.62 | 3.71 | 2.86 |
| PBE-D2+UJ | 3.57 | 3.67 | 2.90 |
| PBE-D3+UJ | 3.61 | 3.71 | 2.88 |
| optB86b-vdw+UJ | 3.56 | 3.60 | 2.88 |
| optB88-vdw+UJ | 3.59 | 3.62 | 2.88 |
| SCAN-rVV10+UJ | 3.63 | 4.00 | 2.83 |

Table 4 Lattice Constant, magnetic moment per Mn, layer thickness $d_1$ and interlayer spacing $d_2$ of bilayer MnSe$_2$ with various exchange-correlation functionals.

| Functional | Lattice Constant(Å) | Mag.Mn($\mu_B$) | $d_1$(Å) | $d_2$(Å) |
|---|---|---|---|---|
| PBE-w/o UJ | 3.48 | 2.89 | 2.75 | 3.26 |
| PBE+UJ | 3.65 | 3.78 | 2.85 | 3.00 |
| PBE-D2+UJ | 3.59 | 3.74 | 2.91 | 2.91 |
| PBE-D3+UJ | 3.63 | 3.78 | 2.86 | 2.71 |
| optB86b-vdw+UJ | 3.59 | 3.70 | 2.88 | 2.70 |
| optB88-vdw+UJ | 3.62 | 3.72 | 2.88 | 2.81 |
| SCAN-rVV10+UJ | 3.67 | 4.07 | 2.80 | 2.78 |

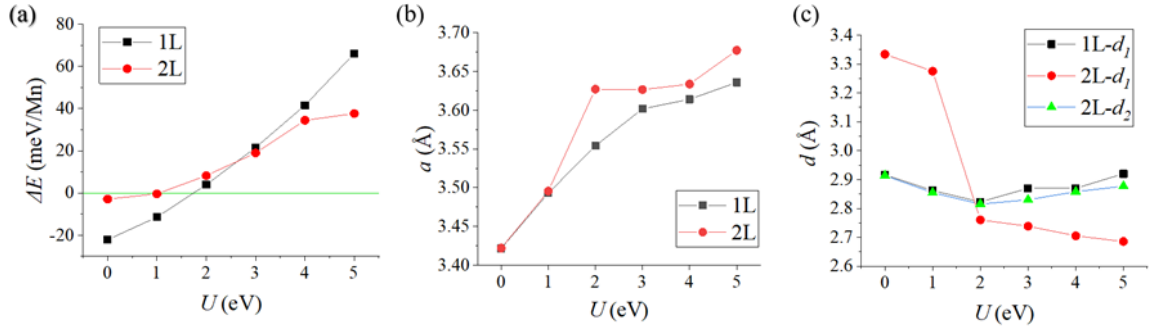

Fig. 5 (a) Energy differences between FM and AFM configurations ($\Delta E=E_{AFM}-E_{FM}$) (b) lattice constant $a$, and (c) thickness $d_1$ and interlayer spacing $d_2$ with respect to different $U$ values. Here, we choose the AFM configuration with lowest energy among all AFM orders in mono- or bilayer MnSe$_2$ as a representative. By using the linear response method calculated $U$ value of 4 eV, our calculations indicate the magnetic ground states and geometric structures converges and should be robust in our calculations.

# APPENDIX B: MAGNETIC GROUND STATES AND HEISENBERG EXCHANGE PARAMETERS

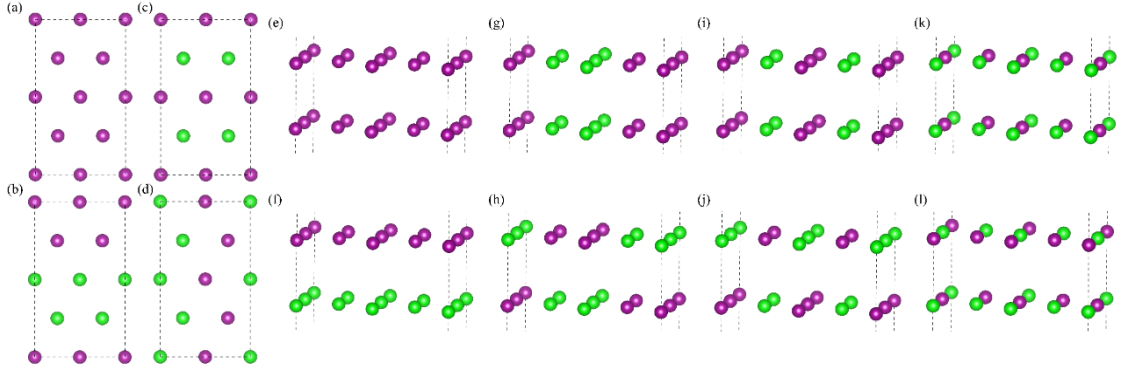

Fig. 6 Magnetic ground states and spin exchange parameters of mono-/bi-layer MnSe$_2$. (a-d) Top views of schematics showing intralayer magnetic orders, including FM (a), AABB (b), ABAB (c) and ZZ (d) in monolayer/bilayer MnSe$_2$, respectively. (e-l) Schematics of eight magnetic orders used for finding the magnetic groundstate and deriving spin-exchange parameters of bilayer MnSe$_2$. Purple and green balls represent two anti-parallel orientations of magnetic moments on Mn atoms, respectively.

Magnetic energies of these magnetic configurations in each magnetic unit cell read as follow:

$$E_e = -\frac{N^2}{4} \times \frac{1}{2}\left(6J_1 + 6J_2 + 6J_3 + \frac{1}{2}J_4 + 3J_5 + 3J_6\right) + E_0$$

$$E_f = -\frac{N^2}{4} \times \frac{1}{2}\left(6J_1 + 6J_2 + 6J_3 - \frac{1}{2}J_4 - 3J_5 - 3J_6\right) + E_0$$

$$E_g = -\frac{N^2}{4} \times \frac{1}{2}\left(2J_1 - 2J_2 - 2J_3 + \frac{1}{2}J_4 + J_5 - J_6\right) + E_0$$

$$E_h = -\frac{N^2}{4} \times \frac{1}{2}\left(2J_1 - 2J_2 - 2J_3 - \frac{1}{2}J_4 - J_5 + J_6\right) + E_0$$

$$E_i = -\frac{N^2}{4} \times \frac{1}{2}\left(-2J_1 - 2J_2 + 6J_3 + \frac{1}{2}J_4 - J_5 - J_6\right) + E_0$$

$$E_j = -\frac{N^2}{4} \times \frac{1}{2}\left(-2J_1 - 2J_2 + 6J_3 - \frac{1}{2}J_4 + J_5 + J_6\right) + E_0$$

$$E_k = -\frac{N^2}{4} \times \frac{1}{2}\left(-2J_1 + 2J_2 - 2J_3 + \frac{1}{2}J_4 - J_5 + J_6\right) + E_0$$

$$E_l = -\frac{N^2}{4} \times \frac{1}{2}\left(-2J_1 + 2J_2 - 2J_3 + \frac{1}{2}J_4 + J_5 - 3J_6\right) + E_0$$

where $N$ represents the unpaired spins on each Mn atom, which is treated as 3 in our exchange parameter calculations, and $E_0$ represents the energy that is independent of the magnetic configuration.

Table 5 Relative total energies of mono-/bi-layer MnSe$_2$ in different magnetic configurations shown in Fig. 6. The term "Mag. Config." is the abbreviation of magnetic configuration. We set the total energy of the FM(-FM) configuration as the reference zero.

| Layer Number | Mag. Config. | $\Delta E$ (meV/Mn) |
|---|---|---|
| Monolayer | FM | 0.0 |
| | AABB | 42.0 |
| | ABAB | 77.7 |
| | ZZ | 70.3 |
| bilayer | FM-FM | 0.0 |
| | FM-AFM | 31.4 |
| | AABB-FM | 62.1 |
| | AABB-AFM | 72.4 |
| | ABAB-FM | 100.4 |
| | ABAB-AFM | 94.9 |
| | ZZ-FM | 102.4 |
| | ZZ-AFM | 108.1 |

## APPENDIX C: ANISOTROPIC SPIN EXCHANGE PARAMETERS AND KITAEV INTERACTIONS

To investigate the effects of non-collinear spin exchange and Kitaev interactions on magnetic anisotropy, we considered an anisotropic Hamiltonian containing both anisotropic spin exchange coupling (SEC) and single ion anisotropy(SIA) terms:

$$H = H_{EX} + H_{SIA} = -\frac{1}{2}\left[\sum_{i \neq j} \boldsymbol{S}_i \cdot \boldsymbol{J}_{ij} \cdot \boldsymbol{S}_j + 2\sum_i \boldsymbol{S}_i \cdot \boldsymbol{A} \cdot \boldsymbol{S}_i\right]$$

where $\boldsymbol{J}_{ij}$ is the anisotropic Heisenberg exchange parameter matrix and $\boldsymbol{A}$ is a vector representing the single-ion anisotropy. We assume that $J_{ij}$ is symmetric and $A_y=0$. Following the methodology outlined in our previous research [43,57], we derived the parameters and transformed the $J_{ij}$ matrix into the coordinate system $\{\alpha\beta\gamma\}$, corresponding to the Mn-Se bond directions. The resulting $J_{ij}$ matrix is:

$$J_{ij} = \begin{pmatrix} 8.73 & -0.02 & 0.04 \\ -0.02 & 8.76 & 0.05 \\ 0.04 & 0.05 & 8.66 \end{pmatrix}$$

for the monolayer, and:

$$J_{ij} = \begin{pmatrix} 11.47 & 0.02 & 0.08 \\ 0.02 & 11.51 & 0.02 \\ 0.08 & 0.02 & 11.47 \end{pmatrix}$$

for the bilayer.

Because the nondiagonal elements are negligible in $J_{ij}$, we can express the $H_{EX}$ as:

$$H_{EX} = -\frac{1}{2}\sum_{i \neq j}(\lambda_\alpha S_i^\alpha S_j^\alpha + \lambda_\beta S_i^\beta S_j^\beta + \lambda_\gamma S_i^\gamma S_j^\gamma) = -\frac{1}{2}\sum_{i \neq j}(J\mathbf{S}_i \cdot \mathbf{S}_j + K S_i^\gamma S_j^\gamma)$$

where $J = (\lambda_\alpha + \lambda_\beta)/2$ represent the isotropic nearest-neighbor exchange coupling in the $\alpha\beta$-plane and $K = \lambda_\alpha - J$ is the Kitaev anisotropic nearest--neighbor exchange coupling parameter. Using these formulations, we can calculate the values of $J$ and $K$ for both monolayer and bilayer, as listed in Table 6.

Table 6 $J$ and $K$ in monolayer and bilayer MnSe$_2$.

| Layer Number | $J$ (meV/Mn) | $K$ (meV/Mn) |
| --- | --- | --- |
| 1L | 8.75 | 0.11 |
| 2L | 11.47 | 0.04 |

## APPENDIX D: STACKING ORDER TUNABLE MAGNETIC GROUND STATES AND EASY-AXIS DIRECTIONS

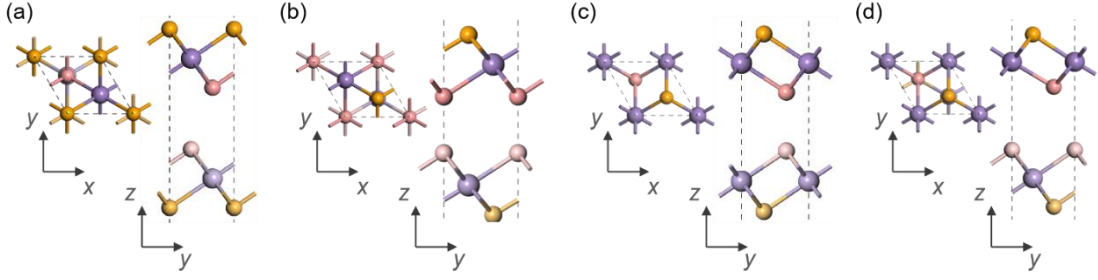

Fig 7 Structure models of bilayer MnSe$_2$ in AB(a), AC(b), AA$^R$(c) and AB$^R$(d) stacking orders.

We consider several common stacking orders in TMD materials [28]. As shown in Table 7, We found AA stracking is most energetically favored among all considered stacking orders with an energy difference of at least 4 meV/Mn. Stacking orders can effectively mudulated the interlayer magneitc groud states (AC and AA$^R$ stacking) and easy axis direction (AB stacking).

Table 7. Magnetic properties and relative energy of bilayer MnSe$_2$ of stacking orders.

| Stacking | Magnetic Ground State | $E$ (meV/Mn) | Easy Axis | MAE (meV/Mn) |
|---|---|---|---|---|
| AA | FM-FM | 0 | Out of Plane | 0.6 |
| AB | FM-FM | 18 | In Plane | 0.1 |
| AC | FM-AFM | 11 | Out of Plane | 2.9 |
| AA$^R$ | FM-AFM | 20 | Out of Plane | 1.5 |
| AB$^R$(same as AC$^R$) | FM-FM | 4 | Out of Plane | 0.7 |

### APPENDIX E: K-MESH DEPENDENCE OF MAE CALCULATIONS

The magnetic anisotropy energy (MAE) is highly sensitive to the choice of k-mesh. To ensure the accuracy of our calculations, we performed extensive convergence tests on the k-mesh using the more accurate tetrahedron method with Blöchl corrections for total energy calculations. As shown in Fig. 8, $E_x$-$E_z$ converges to 0.01 meV/Mn for both the monolayer and bilayer when the *k*-mesh is set to 29×29. For atomically and orbitally decomposed MAE contributions, the *k*-point convergence is even better. These results ensure the robustness of the conclusions presented in our study.

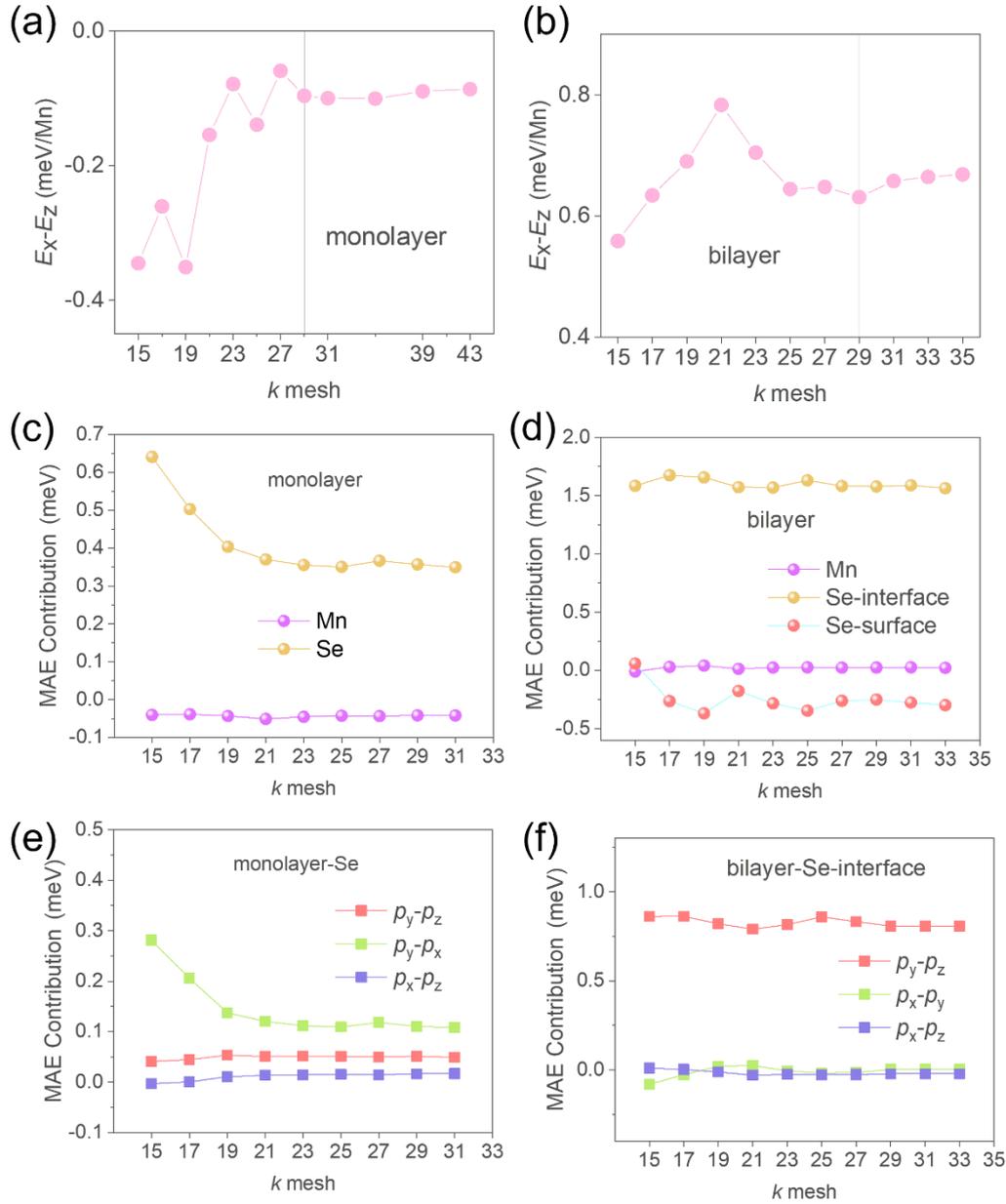

Fig.8 (a-b) Energy difference with respect to $k$-mesh in monolayer (a) and bilayer (b), where $E_x$, $E_z$, represent the energy of the magnetic moment along the directions of $x$ and $z$ axis. (c-f) K-mesh dependence of atomically and orbitally decomposed MAE contributions.

## APPENDIX F: ORBITAL-RESOLVED MAE CONTRIBUTIONS

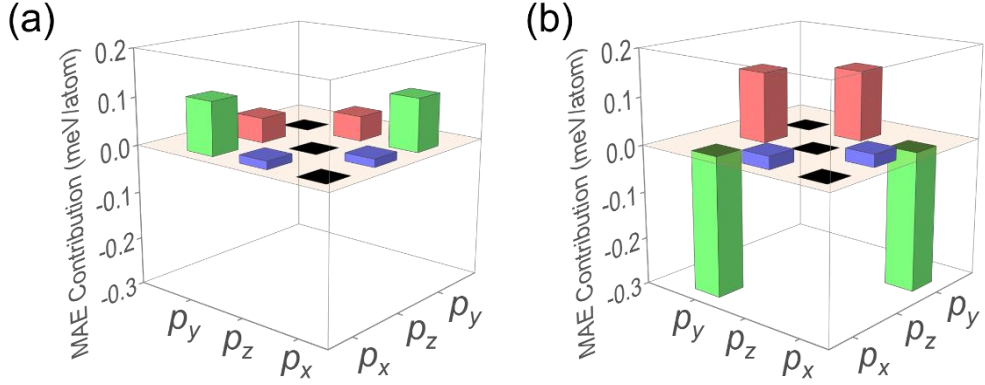

Fig. 9 Orbital-resolved MAE contribution of Se in monolayer(a) and Se-surface in bilayer(b).

## APPENDIX G: ELECTRONIC STRUCTURES WITH ORBITAL DECOMPOSITION

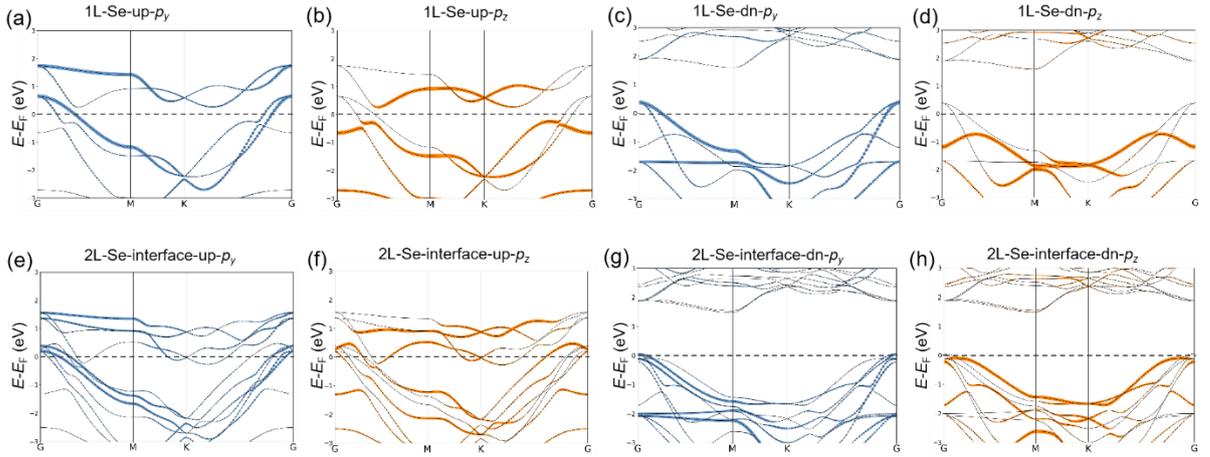

Fig. 10 Band structures of monolayer(a-d) and bilayer(e-h) MnSe$_2$. The (interfacial) Se $p_z$ and $p_y$ orbitals are mapped with different colors: orange, $p_z$; blue, $p_y$.

The states and their contributions to MAE can be related through the second-order perturbation theory [55]:

$$MAE = \sum_{u,o} \frac{|\langle u|H_{SOC}(x)|o\rangle|^2 - |\langle u|H_{SOC}(z)|o\rangle|^2}{E_o - E_u}$$

where $o$ and $u$ correspond to occupied and unoccupied states, respectively. The energy differences between occupied and unoccupied states ($E_o$-$E_u$) are in the denominator, indicating that states near the Fermi level have a greater influence on the MAE compared to those further away.

## APPENDIX H: BAND STRUCTURES WITH THE INCLUSION OF SPIN-ORBIT COUPLING

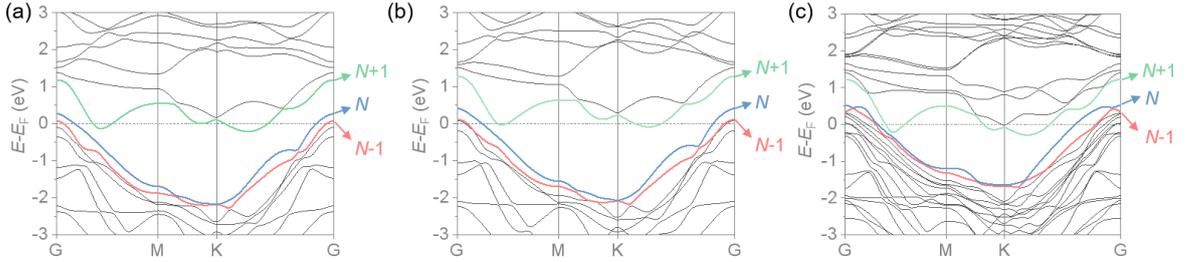

Fig. 11 (a-b) Band structures of monolayer(a,b) and (c)bilayer MnSe$_2$ with SOC, while the magnetic moment is along the easy axis of the monolayer in (a) and along the $z$-axis in (b,c). The symbols $N$-1, $N$, $N$+1 denote bands with respective colors.

## APPENDIX I: EFFECTS OF DOPING, STRAIN, AND STACKING ORDERS ON EASY AXIS DRECTION

Table 8 Lattice constants, easy axis direction and magnetic anisotropic energies of monolayer MnSe$_2$ upon electron/hole doping. FM remains the ground state at different doping concentrations.

| Electron doping(e/Se) | Lattice Constant (Å) | Easy Axis | | |
|---|---|---|---|---|
| | | $\theta(°)$ | $\phi(°)$ | $E_x - E_{ea}$ (meV/Mn) |
| -0.1 | 3.58 | 90 | 90 | 0.01 |
| -0.05 | 3.60 | 90 | 116 | 0.01 |
| -0.0375 | 3.61 | 89 | 116 | 0.01 |
| -0.025 | 3.61 | 90 | 112 | 0.01 |
| 0 | 3.61 | 68 | 111 | 0.01 |
| 0.05 | 3.61 | 17 | 96 | 0.29 |
| 0.075 | 3.61 | 0 | 90 | 0.74 |
| 0.1 | 3.63 | 0 | 90 | 0.81 |

Table 9 Lattice constants, easy axis directions and magnetic anisotropic energies of monolayer MnSe$_2$ under biaxial strain. FM remains the ground state under strain.

| Strain (%) | Lattice Constant (Å) | Easy Axis | | |
|---|---|---|---|---|
| | | $\theta$ (°) | $\phi$ (°) | $E_x - E_{ea}$ (meV/Mn) |
| -3 | 3.50 | 15 | 34 | 0.13 |
| -2 | 3.54 | 0 | 90 | 0.29 |
| -1 | 3.57 | 90 | 114 | 0.01 |
| 0 | 3.61 | 68 | 111 | 0.01 |
| 1 | 3.65 | 16 | 109 | 0.31 |
| 2 | 3.68 | 0 | 90 | 0.93 |
| 3 | 3.72 | 0 | 90 | 0.35 |

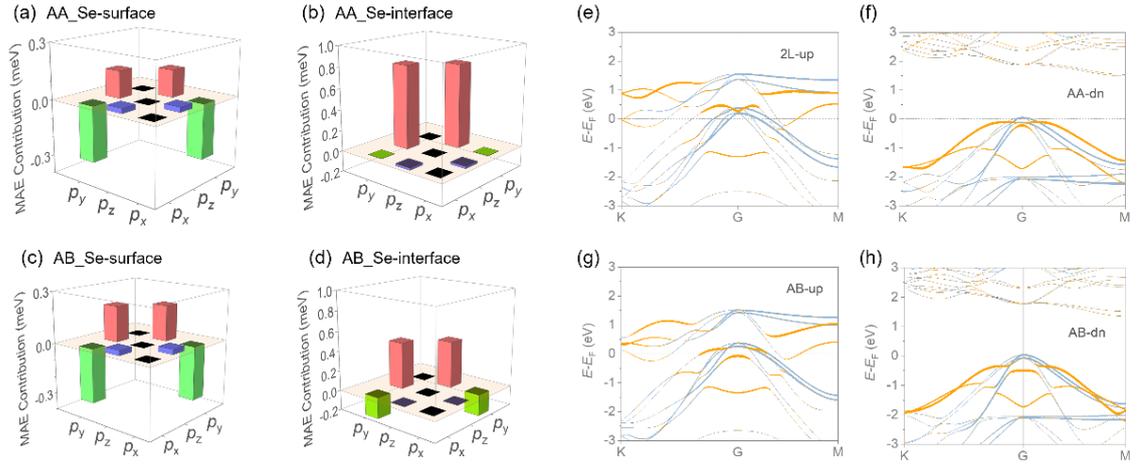

Fig. 12 Orbital-resolved MAE contribution of Se in AA(a-b) and AB(c-d) stacking. (e-h) Band structures of AA(e-f) and AB(g-h) stacked bilayer $MnSe_2$. The selected $p_z$ and all $p_y$ orbitals of (interfacial) Se are mapped with different colors: orange, $p_z$; light blue, $p_y$. The Fermi level is marked using grey dot dash line. From the orbital-resolved MAE in (a-d), we found that the most notable difference between AA and AB stacking lies in the weakened interaction between $p_z$ and $p_y$ states in AB stacking. This change arises from the shift of $p_z$ states in AB stacked bilayer.